\documentclass{spie}  
\usepackage{graphicx}
\usepackage{subfigure}

\title{Tests of achromatic phase shifters performed 
on the S{\large\bf YNAPSE} test bench: a progress report} 

\author{Pavel Gabor\supit{a},
Peter A. Schuller\supit{a},
Bruno Chazelas\supit{b},
Michel Decaudin\supit{a},
Alain Lab\`eque\supit{a},
Philippe Duret\supit{a},
Yves Rabbia\supit{c},
Ralf Launhardt\supit{d},
Jean Gay\supit{c},
Zoran Sodnik\supit{e},
Marc Barillot\supit{f},
Frank Brachet\supit{g},
Thomas Laurent\supit{h},
Sophie Jacquinod\supit{a},
Denis Vandormael\supit{i},
J\'er\^ome Loicq\supit{i},
Dimitri Mawet\supit{j},
Marc Ollivier\supit{a}, 
Alain L\'eger\supit{a}
\skiplinehalf
\supit{a}Institut d'Astrophysique Spatiale, Univ. Paris-Sud 11, Orsay, France; \\
\supit{b}Observatoire de Gen\`eve, Gen\`eve, Switzerland; \\
\supit{c}Observatoire de la C\^ote d'Azur, Nice, France; \\
\supit{d}Max-Planck-Institut f\"ur Astronomie, Heidelberg, Germany; \\
\supit{e}European Space Agency, Nordwijk, Netherlands; \\
\supit{f}Thales Alenia Space, Cannes La Bocca, France; \\
\supit{g}Centre National d'Etudes Spatiales, Toulouse, France; \\
\supit{h}Univ. Li\`ege, Li\`ege, Belgium; \\
\supit{i}Centre Spatial de Li\`ege, Li\`ege, Belgium; \\
\supit{j}Univ. Li\`ege, Institut d'Astrophysique et de G\'eophysique, Li\`ege, Belgium 
}

\authorinfo{Copyright 2008 Society of Photo-Optical Instrumentation Engineers 
(SPIE). This paper was published in Markus Schöller, William C. Danchi, Françoise Delplancke (eds.), {\em Optical and Infrared Interferometry,} 
SPIE Proceedings series, Vol. 7013, paper 70134O (July 28, 2008),
and is made available
as an electronic reprint with permission of SPIE. 
One print or electronic copy may be made for personal use only.
Systematic or multiple reproduction, distribution to multiple 
locations via electronic or other means, duplication of any material in
this paper for a fee or for commercial purposes, or modification of 
the content of the paper are prohibited.
Corresponding author: p.gabor@jesuit.cz}

\begin{document} 
  \maketitle 

\begin{abstract}
The achromatic phase shifter (APS) is a component of the Bracewell 
nulling interferometer studied in preparation for future space missions 
(viz. \emph{Darwin}/TPF-I) focusing on spectroscopic study 
of Earth-like exo-planets. Several 
possible designs of such an optical subsystem exist. Four approaches were 
selected for further study. Thales Alenia Space
developed a dielectric prism APS.
A focus crossing 
APS prototype was developed by the OCA, Nice, France. A field reversal APS 
prototype was prepared by the MPIA in Heidelberg, Germany. 
Centre Spatial de Li\`ege develops a concept based on Fresnel's rhombs.
This paper presents a progress report on the current work 
aiming at evaluating these prototypes
on the {\sc Synapse} test bench
at the Institut d'Astrophysique Spatiale in Orsay, France.  
\end{abstract}



\section{NULLING INTERFEROMETRY AND ACHROMATIC PHASE SHIFTERS}
\label{sec:intro}  

Nulling interferometry, based on the concept 
suggested by Bracewell\cite{Art_Bracewell_1978} in 1978, 
is one of the methods of future exoplanet research.  
It has been studied for the mission 
\emph{Darwin}\cite{2003toed.conf...41K} 
proposed to ESA as well as for the
mission TPF-I (Terrestrial Planet Finder - 
Interferometry)\cite{TPF_Coulter} proposed to NASA. 

The basic performance parameter of a nulling interferometer is the
``nulling ratio'' (sometimes also referred to as ``stellar leakage'' 
although this term refers more properly to stray starlight due to 
the fact a star is not a point source) 
$nl(\lambda,t)=I_{\min}/I_{\max}$
where $I_{\min}$ and $I_{\max}$ stand for the intensity of the on-axis
dark fringe and of the off-axis bright fringe, respectively. 

The goal of \emph{Darwin}/TPF-I is to perform infra-red 
($6-18\,\mu\mathrm{m}$)
imagery of extrasolar planetary systems and 
spectroscopic observations of exoplanetary atmospheres 
in view of evaluating the presence of biomarkers. 
In order to reach this goal, the required nulling performance has 
been estimated\cite{Variability_Noise} as
$\left\langle nl\right\rangle (\lambda)
=1.0\,10^{-5}\,\left(\lambda/7\,\mathrm{\mu m}\right)^{3.37},$
with a level of long-term stability expressed as
$\sigma_{\left\langle nl\right\rangle }
(7\,\mathrm{\mu m},10\,\mathrm{days})\leq 3\,10^{-9}.$

For a given geometric difference $\delta$ between the two optical
paths, the corresponding phase difference $\Delta\varphi$ 
can be expressed as $\Delta\varphi=2\pi\delta/\lambda$
where $\lambda$ is the wavelength. The phase shift therefore
depends on the wavelength $\Delta\varphi=\Delta\varphi(\lambda).$
In order to obtain a wavelength-independent phase shift
of $\pi$ at 0 OPD (optical path difference), an achromatic 
phase shifter (APS) has to be introduced into the setup. 
(An alternative approach, using an ``adaptive nuller'', 
has been studied at the JPL\cite{Proc_Peters_2006}.)

Bracewell's nulling interferometer includes an APS as an optical
means of eliminating the on-axis flux (stellar light), thus facilitating
the detection of the off-axis flux (containing light
from extrasolar planets).
From the very inception, \emph{Darwin}/TPF-I have opted for
a Bracewell-based design, and therefore need an effective APS.

Many different concepts of achromatic phase shifters are available
in the literature.\cite{Report_Rabbia_2001,Proc_Rabbia_2000} 
Their merits need to
be evaluated quantitatively.
The \emph{Darwin} collaboration studied ten APS
concepts\cite{ESA_APS}, promising $nl<10^{-6}$. The spectral range
was that of $6 - 20\,\mu\mathrm{m}$, with $6 - 18\,\mu\mathrm{m}$ 
mandatory, the
extension to $18 - 20\,\mu\mathrm{m}$ priority number 2, and that to 
$4 - 6\,\mu\mathrm{m}$
priority number 3. Four concepts were selected for further study:

\paragraph{1. Dispersive prisms }

\begin{figure}
\begin{center}
\subfigure[Dielectric prisms]{\includegraphics[width=0.49\textwidth]{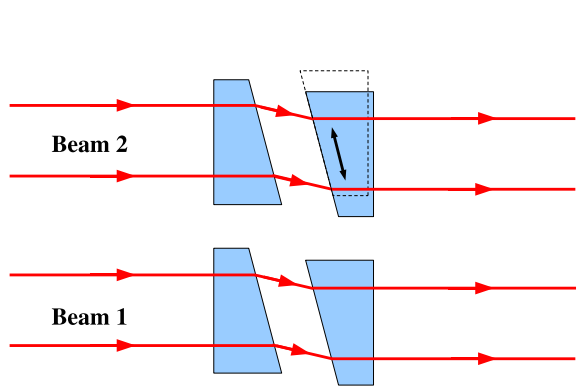}}
\subfigure[Field reversal]{\includegraphics[width=0.49\textwidth]{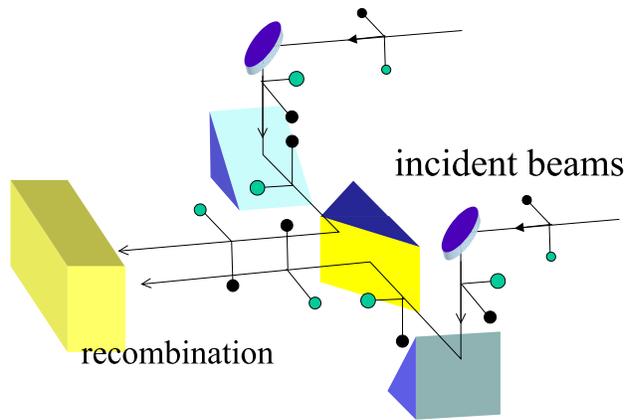}}

\subfigure[Focus crossing]{\includegraphics[width=0.49\textwidth]{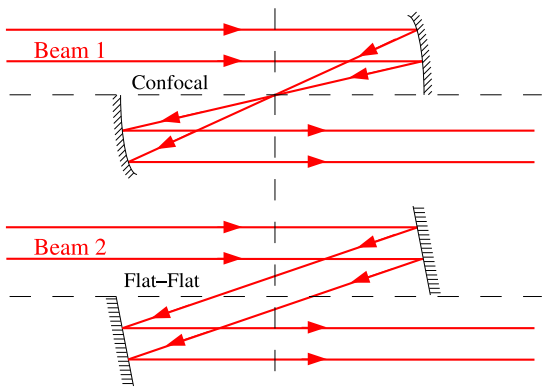}}
\subfigure[Fresnel's rhombs]{\includegraphics[width=0.49\textwidth]{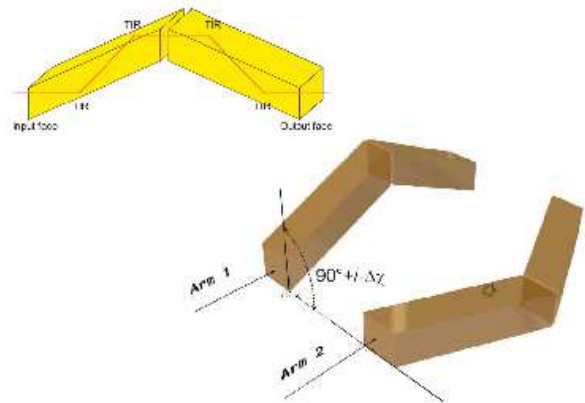}}
\end{center}

\caption{\label{fig:APS_4}The four APS concepts under study 
for \emph{Darwin}.}
\end{figure}

(Fig. \ref{fig:APS_4} a). The principle of this method
consists in introducing dielectric plates into each arm of the interferometer.
Their number, composition and thickness 
are optimised, together with the OPD,
so that the chromaticity of the resulting phase shift in a given
spectral band is below a specified level. As we shall see in the 
description of {\sc Synapse} (Sec. \ref{sec:synapse}), 
a simple APS using two prisms 
of the same material in each arm of the interferometer is an integral part
of the setup.  A model with three prisms of three different materials
per interferometer arm is under development by Thales Alenia Space.

\paragraph{2. Field reversal at reflection}

(Fig. \ref{fig:APS_4} b). From two reflections, arranged in
such a way that s and p polarisation components are successively permuted,
the flip of field vectors provides an achromatic phase shift, as well
as a pupil rotation of $\pi$. Suitably
applied to a couple of interfering beams, a couple of periscope yields
two lightwaves phase shifted by $\pi$ (opposed field vectors). Then
a modified constructive interferometer provides two nulled outputs
by suitably mixing the beams. A prototype APS was developed at 
Max-Planck-Institut f\"ur Astronomie in Heidelberg in collaboration 
with Kayser-Threde GmbH in Munich and the IOF Fraunhofer Institute
for Applied Optics in Jena.\cite{Launhardt2008SPIE}

\paragraph{3. Focus crossing}

(Fig. \ref{fig:APS_4} c). 
The APS FC is based  on the so-called focus-crossing (FC) effect where  
light undergoes  a phase shift of $\pi$ when crossing a focus. The
phenomenon is independent of 
wavelength\cite{Art_Gouy,Art_Boyd}.
The principle of this approach is based on destructive 
interference. It is quite similar to the one of the 
Achromatic Interfero 
Coronagraph\cite{Art_Gay_Rabbia,Art_Baudoz_2000a,Exp_Art_Baudoz_2000b,
Proc_Rabbia_2006}.

APS FC comprises, one bearing two confocal half-parabolic off-axis mirrors, 
the other, two flat mirrors. 
The ``confocal'' module produces 
an achromatic  $\pi$-phase shift, a pupil rotation by $\pi$
and an extra optical path (twice the focal length of a 
paraboloid), while the role of the ``flat-flat'' module 
is to balance this pathlength and to reproduce the
beam geometry of the confocal module.
At beam combination, the resulting 
amplitude is nulled for an on-axis point-like source,
provided the wavefronts are perfectly cophased.
If not, nulling is incomplete and a residual flux remains.

\paragraph{4. Fresnel's rhombs}

(Fig. \ref{fig:APS_4} d). are based on the phase shift 
at total internal reflection. 
It occurs when the incidence at material/air interface
exceeds a limit angle $\vartheta_{\lim}$. A linear polarisation with
azimuth $\pi/4$ at entry, becomes a circular polarisation at exit
(phase shift by $\pi/2$ between s and p components). Two cascaded
rhombs then yield a $\pi$ phase shift. A specific disposition of
two coupled rhombs allows working with any azimuth of the polarisation
at entry (dual polarisation is then allowed).  Centre Spatial de Li\`ege 
has been working on a concept combined with sub-lambda (subwavelength,
i.e., zero order) gratings on the active surfaces 
of the rhombs.\cite{Mawet2006SPIE,Mawet2007OpticsExpress}

\section{THE SYNAPSE TEST BENCH}
\label{sec:synapse}

Since 1995 the Darwin team of the Institut d'astrophysique spatiale,
Orsay, has been developping optical benches for high-contrast interferometry.
The first test bench, operating at 10 $\mu$m, allowed to show that
a monochromatic rejection factor of $10^{3}$ can be reached 
\cite{Art_Ollivier_2001,PhD_Ollivier_1999}.
Nonetheless, as we have already said, the real technical challenge
is to maintain the stability of the nulling performance. Another test bench,
SYmmetric Nuller for Achromatic Phase Shifters Evaluation ({\sc Synapse}),
was designed for the research and development of this 
concept\cite{PhD_Frank_Brachet}.

\subsection{Sources}
\label{subsec:sources}

We have been using three radiation sources:

\paragraph{2000~K Black body source}
consisting of a LaCr hollow cylinder (20 mm diameter, 20 cm length) with 
a 5 mm hole on the side, conducting a current of $\approx 17\,\mathrm{A}$,
emits visible and thermal radiation closely approximating a black body.  
The radiation is filtered using interference filters,
focalised with an off-axis parabola, 
and injected into a single mode optical fiber (SMF; single-mode 
cutoff wavelength 
$\lambda_{c}=1.95\,\mu\mathrm{m}$ or $2.85\,\mu\mathrm{m}$)
providing modal filtering.
In order to avoid thermal and mechanical perturbations of the interferometer,
the source subsystem, as well as the detector subsystem, 
are decoupled from the optical table.

\paragraph{HeNe laser at $3.39\,\mu\mathrm{m}$} with the same filter, 
off-axis parabola and SMF as above.  
With a laser source, the role of the input SMF is 
effectively reduced to bringing the light from the source 
to the mechanically and thermically decoupled interferometer. 

\paragraph{``White laser''} or ``supercontinuum'' source based on a non-linear, 
photonic crystal optical fiber, and producing up to $3\,\mathrm{mW/nm}$
in the range $400 - 2500\,\mathrm{nm}$. After filtering, the source output
is coupled with the same SMF as above, using a spherical mirror.

\subsection{Interferometer}
\label{subsec:interferometer}

\begin{figure}
\includegraphics[width=1\textwidth]{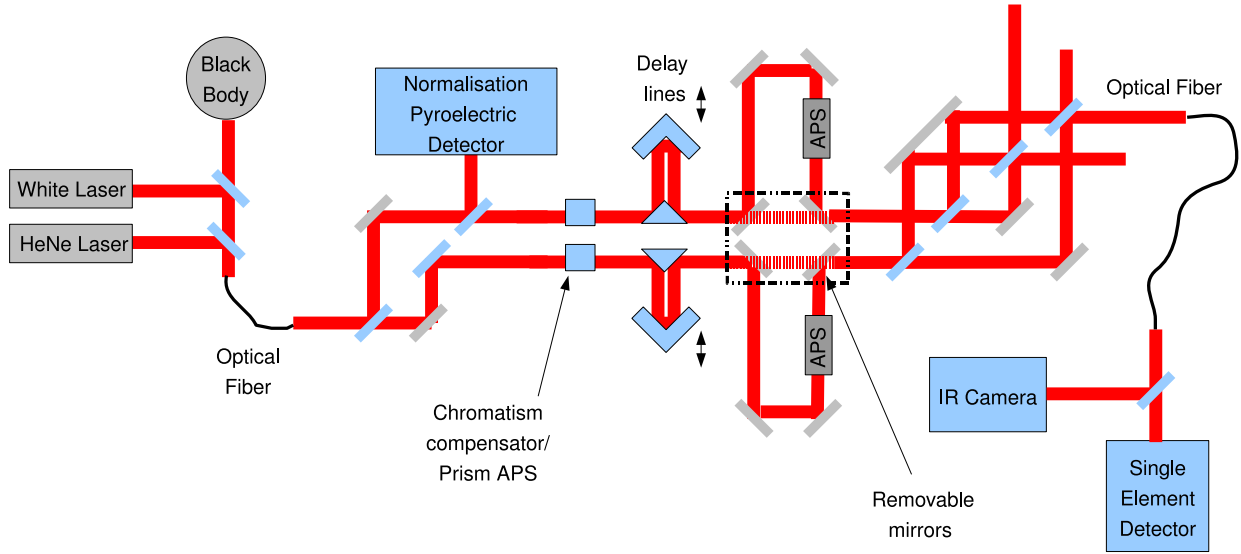}

\caption{\label{fig:synapse-beams}{\sc Synapse} (from left to right): Sources (three
options), SMF, beam-splitter subsystem, CaF$_{2}$ prisms (chromatism compensator and/or APS),
optical switch, APS prototype modules, beam-combiner subsystem, SMF, 
detectors (two options).}
\end{figure}

The interferometer is set up on an optical table 
with passive vibration protection, and
protected from acoustic and thermal perturbations by a plexiglass box.
The optical bench itself  comprises two modified Mach-Zehnder interferometers
serving as beam-splitting and beam-recombination devices \cite{Deph_Art_Serabyn_2001}
(Fig. \ref{fig:synapse-beams}), creating the two identical beams, 
delay lines, and a simple APS 
with two CaF$_{2}$ prisms in each interferometer arm.

Both delay lines, based on rooftop mirrors, are motorised: 
one with a stepper motor
running 25~mm (i.e., optical path of 50~mm) at 
$0.1\,\mu\mathrm{m}$ reproducibility, 
and the other with a piezoelecric actuator running 
$15\,\mu\mathrm{m}$ at a reproducibility better than $0.5\,\mathrm{nm}$.

The beam-splitter and beam-combiner systems use parallel CaF$_{2}$ plates.  Their 
thickness is not guaranteed to be identical to the required accuracy, 
and therefore a compensator needs
to be included in the design.  The bench's intrinsic APS, comprising two CaF$_{2}$ 
prisms in each interferometer arm, fulfills this role.  As an APS, it can introduce
any given phase shift, and reduces the chromaticity in the K or L band so that 
the resulting theoretical nulling performance is better than $\approx 10^{-5}$.

Since the bench (if well aligned) is symmetric by design, 
the flux of the two arms is balanced using simply injection into the exit SMF, 
tweaking the orientation of the folding mirror illuminating the off-axis parabola.
The better the superposition of the two beams, the easier the balancing.

For the tests of APS prototypes, the two identical beams pass 
through the tested modules  before entering the
beam-combiner subsystem.  
In the current configuration, we use a translation stage 
with folding mirrors as
an optical switch to insert the hosted APS FC prototype
into the optical path.

\subsection{Detectors}
\label{subsec:detectors}

In order to normalise the flux as measured at the output of the bench (useful also 
for fine-tuning of beam injection into the SMF), we have installed
a simple pyroelectric detector at one of the redundant outputs of the beam-splitter 
subsystem.

A SMF, placed at the exit of the interferometer,
provides modal filtering.  
Two types of detectors are used:

\paragraph{Single element detector} is a commercial, liquid nitrogen cooled, InSb detector.
Its signal passes through a lock-in amplifier before being digitised.

\paragraph{Array detector} is an astronomical camera system previously used on the CFHT.
A 128x128 pixel InSb array is mounted in a solid nitrogen cooled cryostat.

\section{TESTS AND PRELIMINARY RESULTS}
\label{sec:tests}

\subsection{Evolution}
\label{subsec:evolution}

\begin{figure}
\includegraphics[%
  width=0.5\textwidth,
  keepaspectratio]{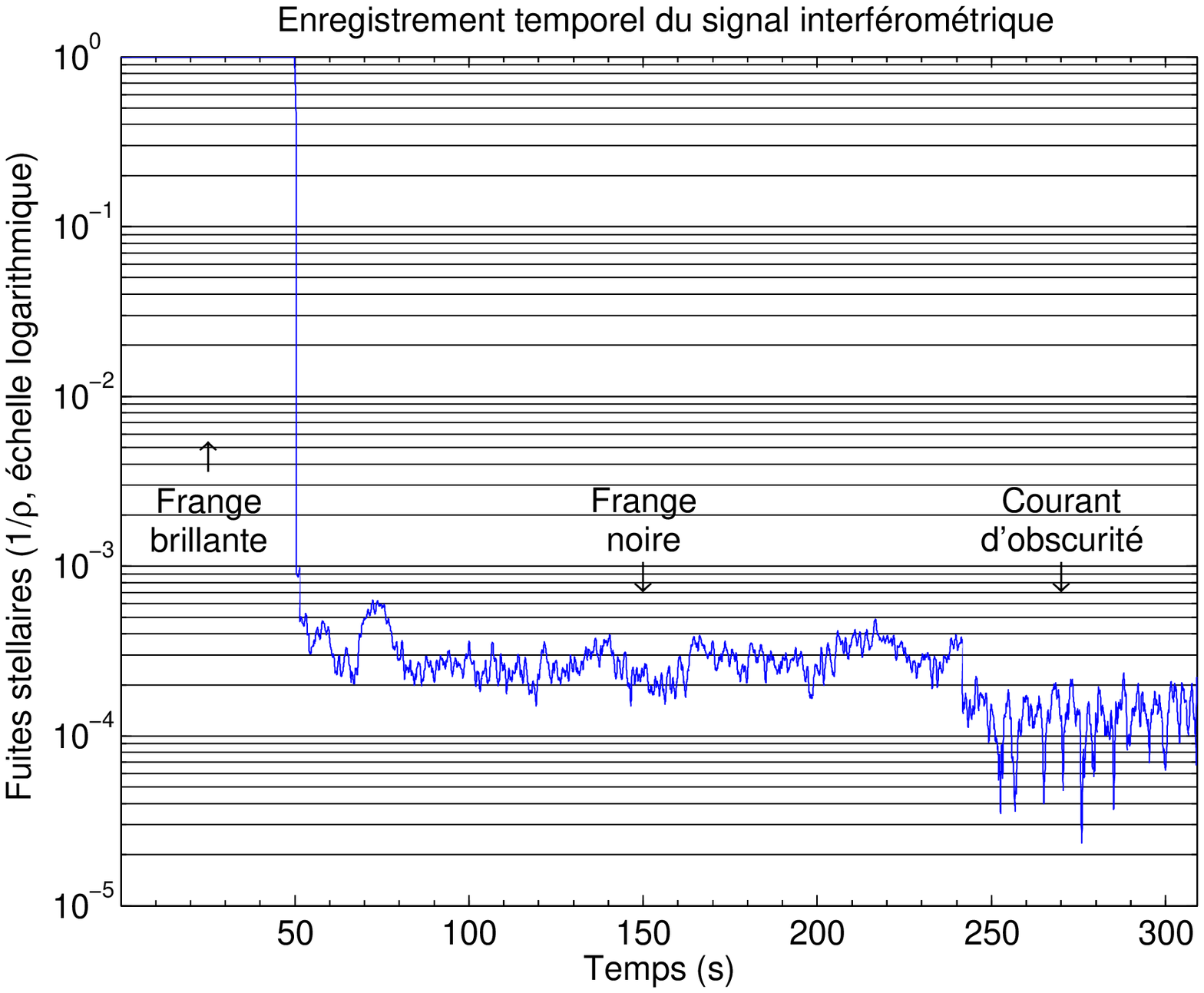}
\includegraphics[%
  width=0.5\textwidth,
  keepaspectratio]{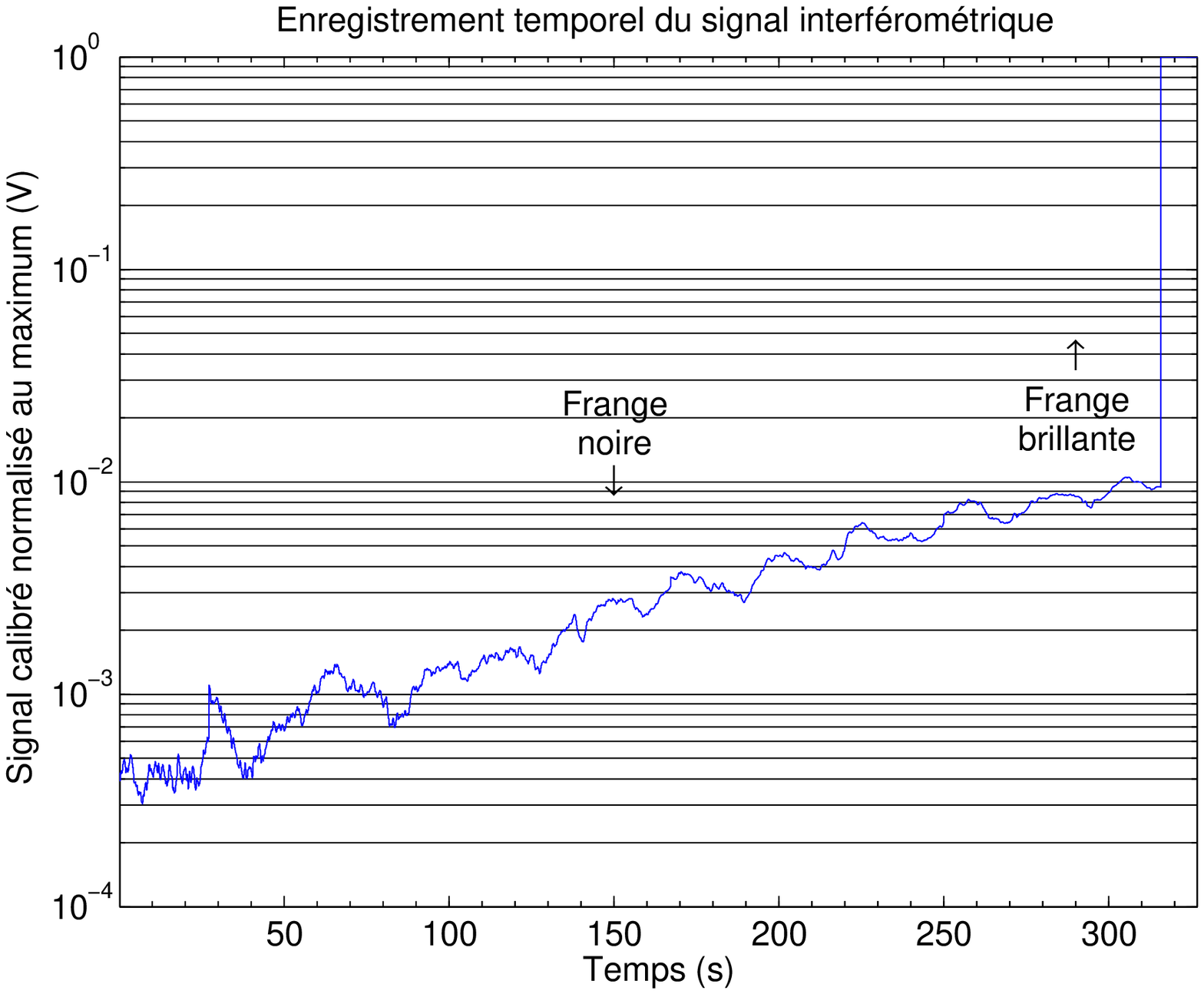}

\caption{\label{fig:frank-acquisition}{\sc Synapse} measurements
in the K spectral band. Left: The
transition between the bright fringe (the first 50 seconds) and the
dark fringe (between 50 s and 240 s) was performed manually by modifying
the length of the delay line. The last part of the curve shows the
dark current (the detector being covered). The mean rejection factor
is about 2.4 times greater than the dark current of the detector.
Right: Same measurement with a significant drift.}
\end{figure}

The work of Frank Brachet \cite{PhD_Frank_Brachet} has documented
two features of the {\sc Synapse} nulling performance in the K band 
($2-2.5\,\mu\mathrm{m}$). The first being the
best value of a nulling ratio of $nl=1.5\,10^{-4}$.
The second parameter observed was the degree of nulling performance
stability. Let us reproduce the results of two typical acquisitions.
The first (Fig. \ref{fig:frank-acquisition}, left) shows an acquisition
of about 200 s with a mean nulling ratio 
$\left\langle nl\right\rangle =2.7\,10^{-4}$
with a considerable standard deviation 
$\sigma_{\left\langle nl\right\rangle }=6\,10^{-5}$.
The second (Fig. \ref{fig:frank-acquisition}, right) is a slightly
longer recording, of about 10 minutes with a 
nulling ratio $\left\langle nl \right\rangle =2.5\,10^{-4}$
during the first 100 seconds 
(with $\sigma_{\left\langle nl \right\rangle }=6\,10^{-5}$),
showing the effects of significant drift, probably due  
to OPD instability.

In order to stabilise the bench, we have introduced a control system 
for OPD\cite{Art_Gabor_2008}, 
implenting dithering (periodic changes) of the OPD close to the optimal value. 
A classical form of this technique has been investigated
\cite{Art_Ollivier_2001}, although this first
experiment was inconclusive. More recently,
Schmidtlin et al.\cite{Schmidtlin2005} 
have demonstrated the potential of dithering
in a sequential
way, obtaining good levels of stabilisation with nulling ratio around 
$8\,10^{-7}$ with a laser diode at 638 nm. It was this work that provided 
our team with decisive inspiration.

Implementing this approach, we have succeeded not only in stabilising 
the nulling performance, but also gained a powerful instrument for bench
optimisation (cf. Sec. \ref{subsec:techniques}).

\begin{figure}
\includegraphics[width=0.5\textwidth]{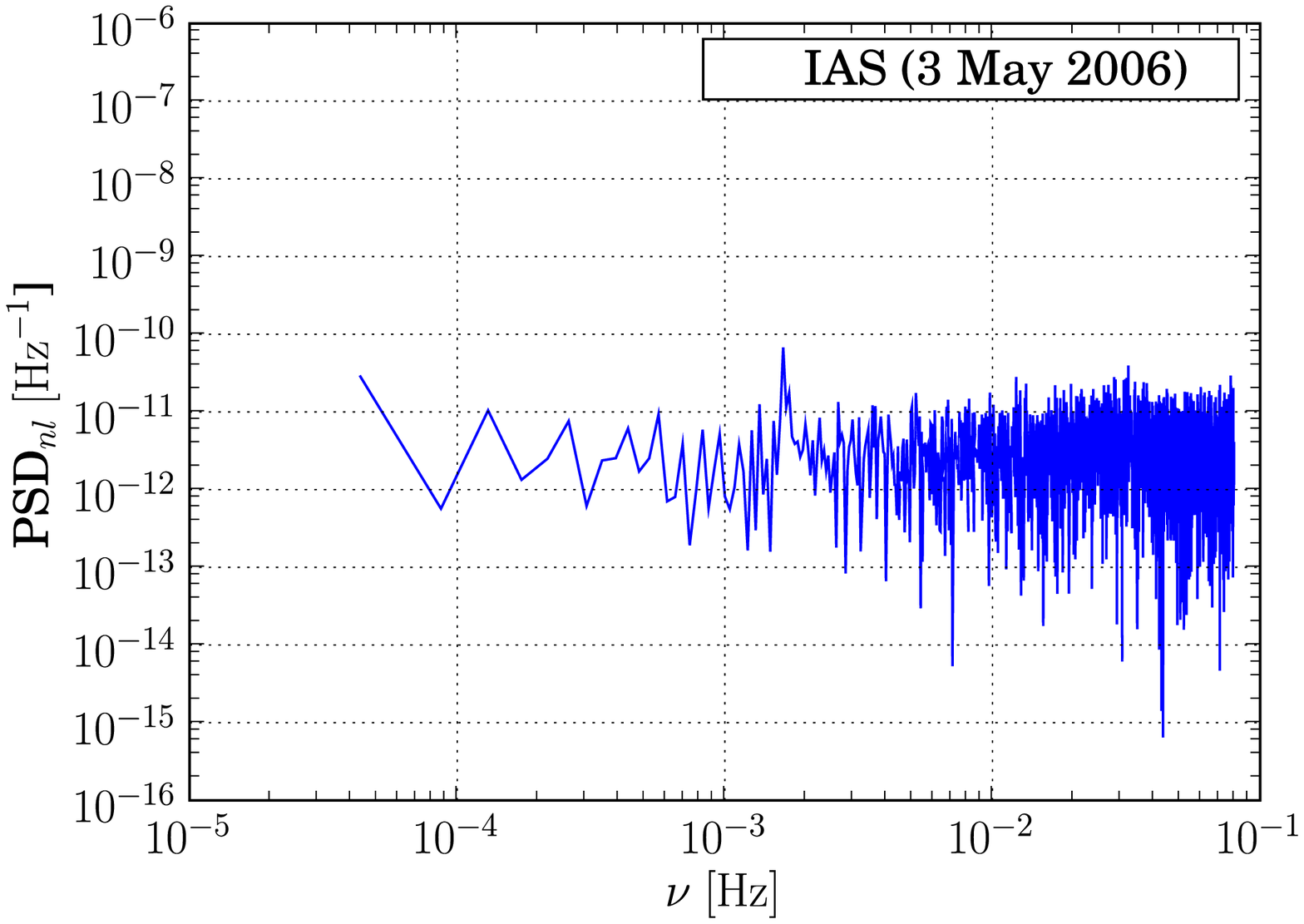}
\includegraphics[width=0.5\textwidth]{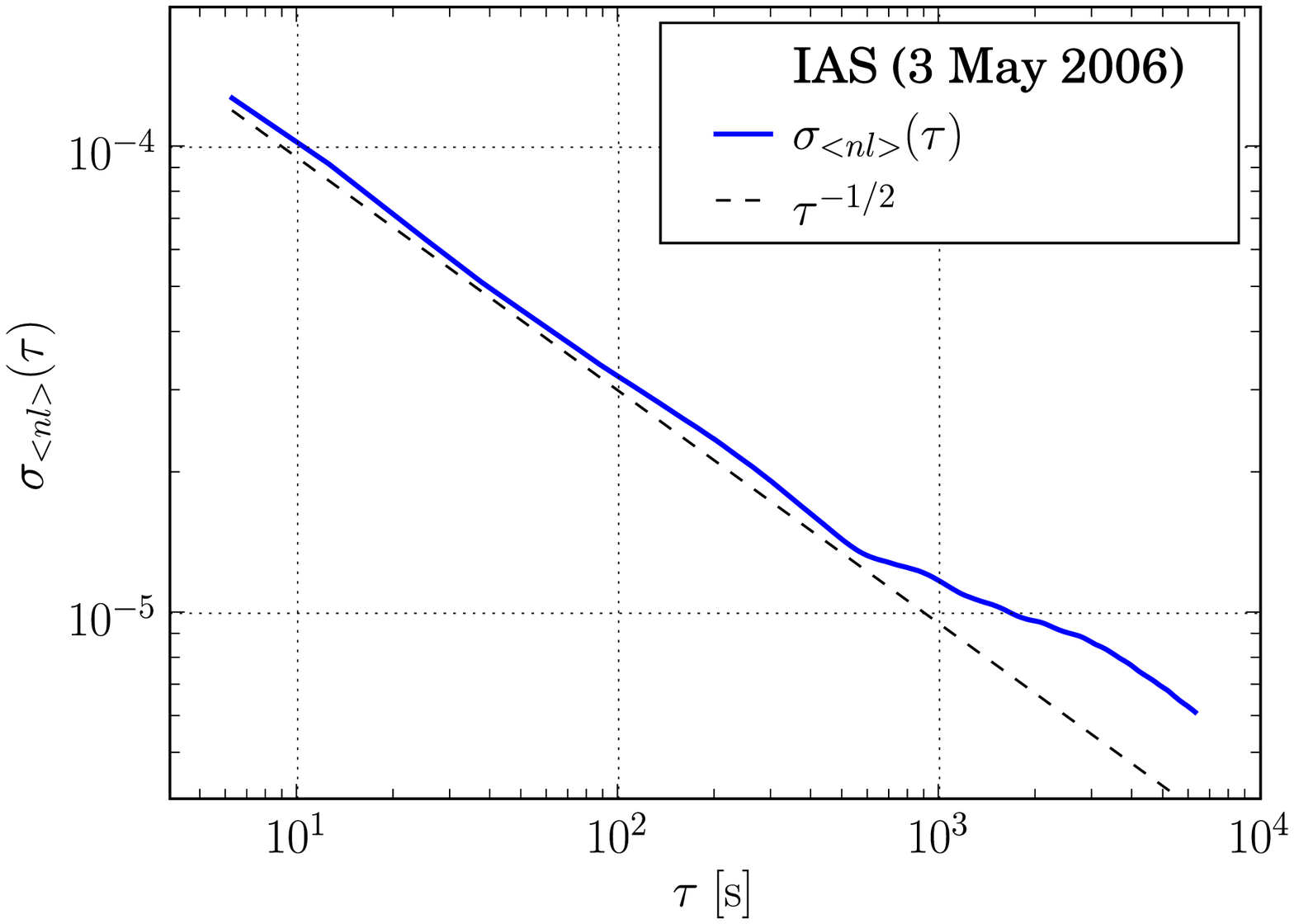}

\caption{\label{fig:psd_sigma}{\sc Synapse} measurements in the 
K spectral band. Left: 
Power spectral density of the nulling function, PSD($nl$).
Note that $1/f$ component is negligible.
Right: Standard deviations of the running average of the
nulling function over the time interval $\tau$, 
$\sigma_{\left\langle nl\right\rangle }(\tau)$
(curve above). A (displaced) $\tau^{-1/2}$ function is shown for
comparison (line below). Note that up to $\tau\approx500$ s, the
experimental curve is consistent with the $\tau^{-1/2}$ behaviour,
with $\sigma_{\left\langle nl\right\rangle }(500\,\mathrm{s})\approx1.5\,10^{-5}$.}

\end{figure}

In the K band, we have obtained $\left\langle nl\right\rangle 
=3\,10^{-4}$ and
$\sigma_{\left\langle nl\right\rangle }
(3\,\mathrm{s}) = 2\,10^{-4}$
at integration times of 3~s, which improves to  
$\sigma_{\left\langle nl\right\rangle }
(2\,\mathrm{hrs}) = 8\,10^{-6}$
after 2 hours of integration.  Since $\sigma$ in Fig. 
\ref{fig:psd_sigma} is 
consistent with $\tau^{-1/2}$ behaviour up to $\tau\approx 500\,\mathrm{s}$, 
with $\sigma_{\left\langle nl\right\rangle }(500\,\mathrm{s})\approx1.5\,10^{-5}$, 
we extrapolate that 
if drifts were kept at bay we would obtain
$\sigma_{\left\langle nl\right\rangle }(10\,\mathrm{days}) 
= 3\,10^{-7}$
after an integration of 10 days, 
still two orders of magnitude short of the goal.

\begin{figure}
\includegraphics[width=0.5\textwidth]{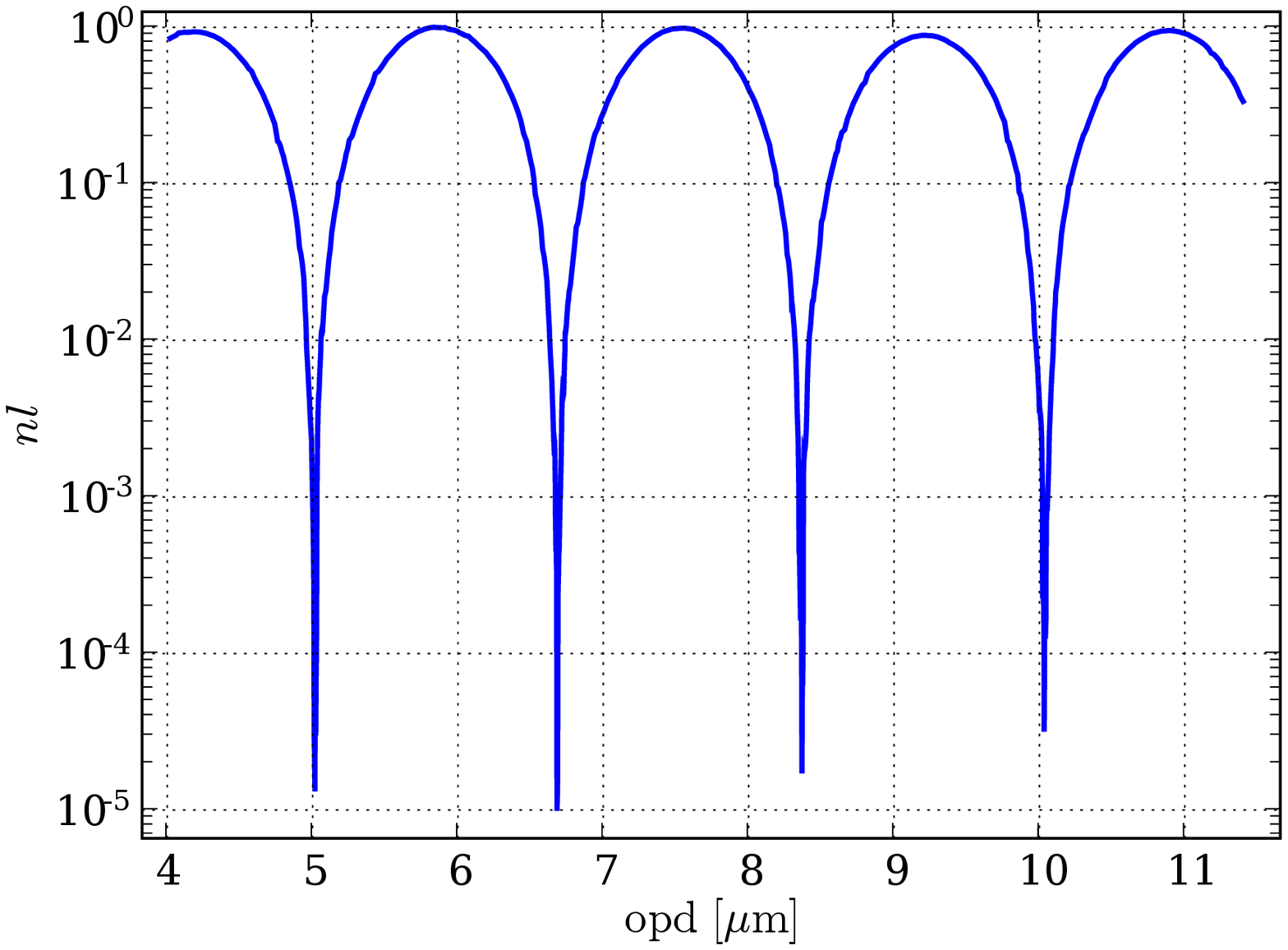}
\includegraphics[width=0.5\textwidth]{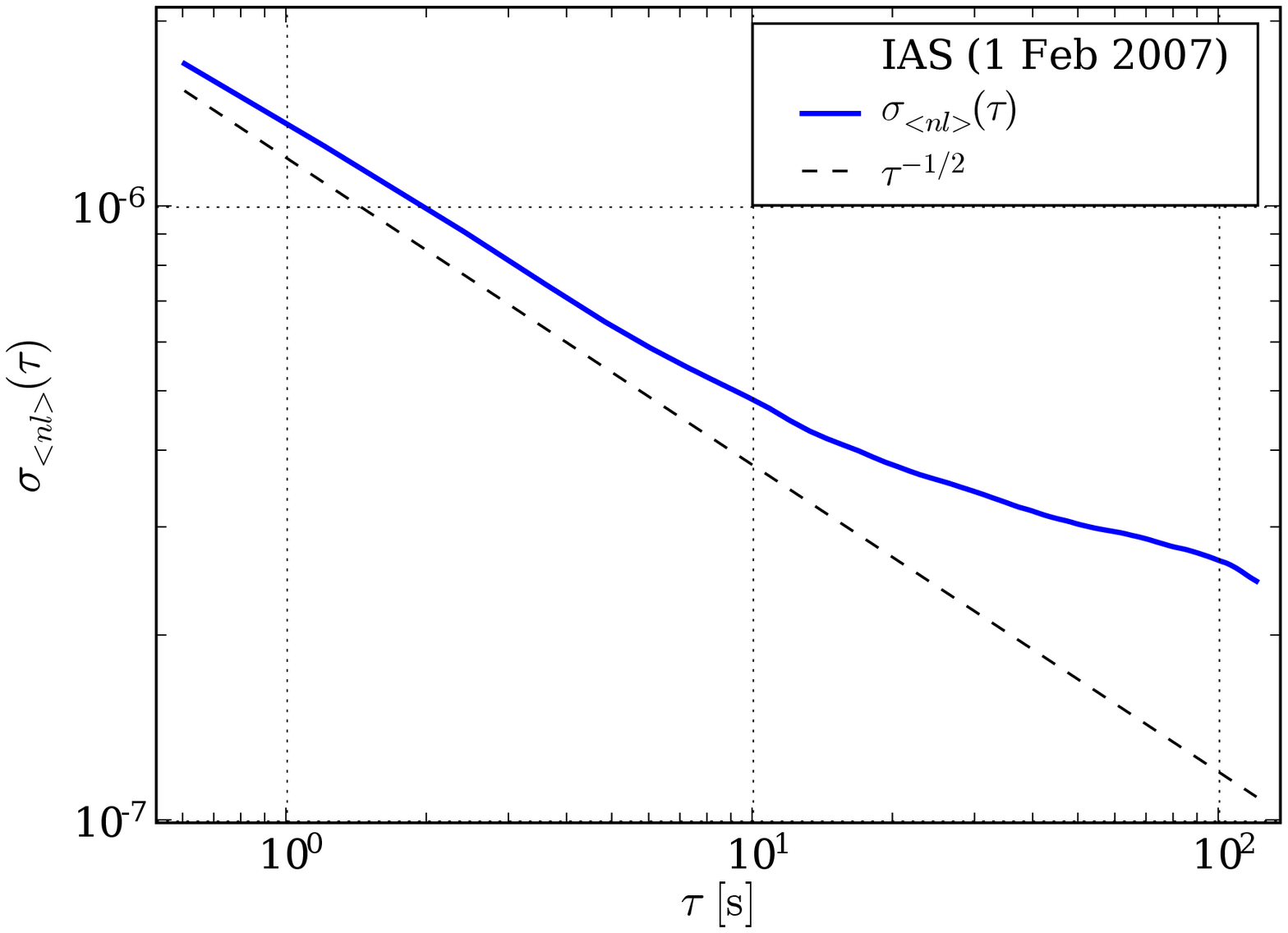}

\caption{\label{fig:null_w_laser}Measurements with 
a HeNe laser at $3.39\,\mu\mathrm{m}$.
Left: Results of an OPD scan in terms of the nulling ratio $nl$.
Right: Results of a 10-minute OPD-stabilised acquisition of the signal 
at the ``bottom'' of the
dark fringe presented in terms of standard deviations 
of the running average of the
nulling function over the time interval $\tau$, 
$\sigma_{\left\langle nl\right\rangle }(\tau)$. 
A (displaced) $\tau^{-1/2}$ function is shown for
comparison (line below).}

\end{figure}

In order to find ways of improving the nulling performance of the bench, we 
started working with monochromatic light, at $3.39\,\mu\mathrm{m}$, 
produced by a HeNe laser.  The first results were limited to 
$\left\langle nl\right\rangle = 6\,10^{-5}$.  Introducing polarisers
has led to 
$\left\langle nl\right\rangle = 1\,10^{-5}$ 
(an OPD scan is shown in Fig.~\ref{fig:null_w_laser}).  
This result
is coherent with our theoretical studies\cite{PhD_Bruno_Chazelas} 
of the effects of polarisation mismatch (primarily due to residual 
alignment errors) on the nulling performance.  

OPD-stabilised long-duration measurements (the OPD dithering was able to
follow the dark fringe over a period $>11\,\mathrm{hrs}$)
have shown, at this level of nulling, instabilities that cannot
be fully controlled with OPD (Fig.~\ref{fig:null_w_laser}, right).
Note that that the system remains at the ``bottom'' of the dark fringe 
(best nulling-ratio values) for typically 10 min although an analysis of 
the noise of the acquired data (plotting standard deviations 
of the running average of the nulling ratio against the width of the
averaging window) exposes
deviations from the $\tau^{-1/2}$, white-noise behaviour.

The next step towards full stabilisation of the bench would have to
involve control of flux mismatch between the interferometer arms. 
In order to perform tests of APS prototypes at $nl=10^{-5}$, however,
OPD stabilisation suffices in providing reliable and reproducible results.

After these encouraging results with monochromatic light 
we concentrated our efforts on improving the level of null in broad band.
From the results of monochromatic measurements we deduced that 
the factor limiting 
nulling levels of the test bench could not be  
thermal and mechanical (vibrations) stability.
We have also seen that improving bench alignement 
(to within $<10\,\mathrm{arcsec}$ 
as opposed to the previous $<3\,\mathrm{arcmin}$) 
may help reduce polarisation mismatch, 
greatly facilitating flux balancing of the two interferometer arms.
An overview of our progress in terms of $nl$ and stability (when
available) is given in Tab.~\ref{tab:overview}.
Re-aligning the bench and accompanying measurements performed 
also enabled us to eliminate other hypotheses, namely, 

\begin{itemize}
\item vignetting, mismatched and/or non-gaussian beam profiles: 
checked in the visible with cameras, and in the IR with pinhole scans;
\item lateral shift and/or angular misalignment of beams injected into 
the exit fiber, relative to each other and/or to the fiber's eigen mode:
checked in the visible to $< 0.1\,\mathrm{mm}$ and $<10\,\mathrm{arcsec}$, 
while calculations tell us that we should be at $nl\approx 10^{-6}$ 
(considering lateral and radial beam chromaticity) 
while the influence of the surface quality of the off-axis parabola
is negligible when the beams are superimposed as specified (otherwise
a flux mismatch of $> 30$ \% between the two beams would be observed);
\item incorrect working point idendification of the bench's intrinsic 
APS (dielectric prisms): different working points, corresponding to 
a combination of OPD and of differential thickness of the dielectric 
in the arms of the interferometer, may give similar fringe patterns 
(because they correspond to phase shifts of $(2n-1)\pi$)
but the optimal (in terms of providing the best
nulling performance) working position is unique; with a degraded null
it is possible to mistake \emph{a} working point for \emph{the} working point; 
we checked this, measuring $nl$ at several such working points,
ranging from $-11\pi$ to $+11\pi$;  
\item spectral mismatch of the two interfering beams: measuring the 
each beam with a spectrometer yielded a mismatch estimate that would
lead to $nl<3.8\:10^{-5}$;
\item asymetry in beam-splitter coatings: assuming differerences in 
thickness of 10 \% we calculated $nl< 10^{-6}$;
\item inhomogeneities in the material of the bench's intrinsic APS: 
checked interferometrically before APS production; we calculated
that such inhomogeneities would lead to $nl<10^{-6}$.
\end{itemize}

\begin{table}
\begin{center}
\begin{tabular}{|l|c|c|c|c|l|}
\hline
 & $\lambda$ & polarisers & APS FC & $nl$ & stability \\
\hline
\hline
Nov '05 & K band & -- & -- & $1.5\,10^{-4}$ & drifts \\
\hline
May '06 & K band & -- & -- & $5\,10^{-4}$ & stable (3 hrs)\\
\hline
Jan '07 & $3.39\,\mu\mathrm{m}$ & -- & -- & $10^{-4}$ & \\
\hline
Jan '07 & $3.39\,\mu\mathrm{m}$ & + & -- & $10^{-5}$ & stable (10 min)\\
\hline
Feb '08 & $3.39\,\mu\mathrm{m}$ & + & -- & $10^{-5}$ & \\
\hline
Feb '08 & $3.39\,\mu\mathrm{m}$ & + & + & $10^{-5}$ & \\
\hline
Feb '08 & K band & -- & + & $4\,10^{-4}$ & \\
\hline
Feb '08 & K band & + & + &  $4\,10^{-4}$ & \\
\hline
\end{tabular}
\end{center}
\caption{\label{tab:overview}An overview of progress on {\sc Synapse} 
in terms of $nl$ and stability (when available).}
\end{table}

\subsection{Techniques}
\label{subsec:techniques}

Several measurement techniques have been studied for {\sc Synapse}:

\begin{itemize}
\item Placing an array detector with a dispersive prism at the exit 
of the bench while scanning OPD, provides a series of low-resolution
spectra.  In an ideal case the resulting pattern would contain a straight
dark fringe.  Its deviation from the ideal form  
yields a measure of phase-shift chromaticity.\cite{PhD_Bruno_Chazelas}  
The chief 
advantages of this approach is that it is inherently independent of
flux mismatch, and that it works even when the signal-to-noise ratio 
does not allow for a direct measurement of $nl$.  
Nonetheless, we have not implemented this approach 
very widely.  One difficulty has to do with OPD drifts, and the 
other is practical: our camera is much more difficult to
use than the single element detector.

\item Another technique which is under study is Fourier transform of OPD scans
with the coarse long line.\cite{PhD_Bruno_Chazelas}  The beams are recombined
in a SMF which also provides modal filtering, and the flux is measured
using a single element detector. The Fourier transform method is 
robust vis-\`a-vis chromatic flux mismatch.  In {\sc Synapse}, 
an improved delay-line actuator
would greatly facilitate the application of this technique.

\item The measurements presented in this paper were obtained following
an experimental protocol which we have been developping for {\sc Synapse}
with the single mode detector (as described in the previous point). Note 
that OPD stabilisation is used not only to reach and maintain the optimal
OPD (and therefore enabling reproducible measurements of $nl$ 
and $\sigma_{nl}$) but also tweaking experiments where the impact 
of small changes in flux balance, CaF$_{2}$ prism position, 
monochromator prism orientation, etc. on the $nl$ and OPD can be monitored.
\begin{itemize}
\item Check position of optical switch: experiment with or without FC APS.
\item Check CaF$_{2}$ prism and long delay-line position.
\item Start optimising injection into the exit fiber by maximising Beam 1 flux
with motorised folding mirrors.
\item Switch to Beam 2.  Remove residual flux mismatch by tweaking 
folding mirror orientation (aiming for $\Delta I/I\approx 0.5\%$).
\item Perform OPD scan with the fine delay line. Find current dark fringe position and the first $nl$ value.
\item Repeat OPD scan, focusing on the dark fringe. Obtain a measure of $nl$.
\item Start OPD-stabilised DAQ, monitoring the flux and OPD position.
\item Perform tweaking experiments: 
\begin{itemize}
\item tweak folding-mirror position in order to monitor/minimise the impact of 
flux mismatch on $nl$;
\item while monitoring its impact on $nl$, tweak CaF$_{2}$ prism position in 
order to fine-tune the prism APS;
\item shift the working wavelength by rotating the monochromator prism, while OPD-stabilisation follows the dark fringe in order to measure corresponding OPD shift.\footnote{This paper does not contain results obtained using this technique.}
\end{itemize}
\end{itemize}
\end{itemize}

\subsection{Prospective}
\label{subsec:perspective}

Pertinent measurements of APS performance on the {\sc Synapse} test bench will be possible
once improved nulling levels are reached.  Identifying the limiting factors 
is therefore our immediate goal.  The tests in progress concentrate on the single-mode
filtering performance of the optical fibers we are using. In a setup 
like {\sc Synapse}, SMF's are crucial in bridging over 
the problem of wavefront bumpiness induced by errors of 
optical surfaces\cite{Art_Mennesson_2002}.
However, the quality of modal filtering provided by a given SMF is
hard to assess\cite{Ksendzov2007}. The work in progress
includes 

\begin{itemize}
\item measurements in a narrow spectral band (16~nm) around $2.3\,\mu\mathrm{m}$, 
\item study of secondary-mode rejection by modifying fiber curvature,
\item measurements in the L band $3-4\,\mu\mathrm{m}$, 
      filtering with a single-mode fiber with $\lambda_{c}=1.95\,\mu\mathrm{m}$,
\item measurements using a prism-based monochromator.
\end{itemize}

\section{DISCUSSION AND CONCLUSION}
\label{sec:disc_n_conc}

The primary objective of this paper was to present an update on our progress 
with the {\sc Synapse} test bench.  Our work has so far 
demonstrated that the setup
reaches a well reproducible monochromatic nulling level $nl=10^{-5}$ 
at $3.39\,\mu\mathrm{m}$ which can be maintained for 10 minutes, thus reaching 
stability of $\sigma_{\left\langle nl\right\rangle }
(3.39\,\mathrm{\mu m},100\,\mathrm{s})\leq 4\:10^{-7}$.
The reproducible nulling level in the K band 
is currently of $nl=3\:10^{-4}$.  We have verified
that it can be maintained for at least several hours, reaching
stability levels of $\sigma_{\left\langle nl\right\rangle }
(2.2\,\mathrm{\mu m},2\:\mathrm{hrs})= 8\:10^{-6}$.

Regarding evaluation of APS prototypes, at the moment, the bench allows for 
work with its ``intrinsic'' APS (a pair of dielectric prisms per branch) and
with the APS FC.  The current primary objective of experiments with the 
APS FC is 
to provide us with more options for our diagnostic tests.
The two APS concepts are in fact very different (one flips the field of view, 
while the other does not, one works only in a set spectral band 
outside of which 
nulling levels deteriorate fast, the other based on a physical 
principle which is 
independent of wavelength), and therefore using both gives 
us more information about the bench itself.

The APS FC was 
initially built to work in the spectral range of $6-18\,\mu\mathrm{m}$ 
(proposed range for \emph{Darwin}/TPF-I), and its
expected nulling performance is  $nl\leq 10^{-6}$,
depending on the bench configuration.  
In the working conditions of {\sc Synapse} the expected 
value is $10^{-5}$. Up to now this performance 
has been demonstrated with laser light at $3.39\,\mu\mathrm{m}$. 
Rejection performance of the APS FC  in the K band 
($2-2.5\,\mu\mathrm{m}$) is to date limited by the performance of the 
{\sc Synapse} bench itself.

Although evaluation of the performance of different APS's may be done 
using fringe dispersion or a monochromator 
(phase-shift measurement as a function of wavelength), clearly the most
representative type of APS estimate calls for measurements with broad-band 
polychromatic light.  We are making progress towards eliminating all possible
limiting factors to {\sc Synapse} broad-band performance.
The work is ongoing and we shall present an update at the SPIE Congress.

\acknowledgements

This work was supported by 
{\em Centre National d'Etudes Spatiales\/} 
({\em action R\&T,\/} n. R-506/SU-002-022) 
and {\em Agence Nationale de la Recherche\/} 
({\em th{\`e}me blanc,\/} n. 5A 0617).


\bibliographystyle{/home/pgabor/texmf/tex/latex/spie/spiebib}
\bibliography{/media/sda5/_science/Darwin_TPFI/biblio/biblio_generale_pg}

\end{document}